**The use of spectral indices in environmental monitoring of smouldering coal-waste dumps**

Anna Abramowicz*, Michał Laska, Ádám Nádudvari, Oimahmad Rahmonov

University of Silesia in Katowice, Faculty of Natural Sciences, Institute of Earth Sciences, Będzińska 60, 41-200 Sosnowiec, Poland

*Corresponding author: Anna Abramowicz, anna.abramowicz@us.edu.pl, phone number: +480323689368

**Keywords:** burning coal heaps; Upper Silesian Coal Basin; thermal analysis; remote sensing; ecological indices

**Abstract**

The study aimed to evaluate the applicability of environmental indices in the monitoring of smouldering coal-waste dumps. A dump located in the Upper Silesian Coal Basin served as the research site for a multi-method analysis combining remote sensing and field-based data. Two UAV survey campaigns were conducted, capturing RGB, infrared, and multispectral imagery. These were supplemented with direct ground measurements of subsurface temperature and detailed vegetation mapping. Additionally, publicly available satellite data from the Landsat and Sentinel missions were analysed. A range of vegetation and fire-related indices (NDVI, SAVI, EVI, BAI, among others) were calculated to identify thermally active zones and assess vegetation conditions within these degraded areas. The results revealed strong seasonal variability in vegetation indices on thermally active sites, with evidence of disrupted vegetation cycles, including winter greening in moderately heated root zones – a pattern indicative of stress and degradation processes. While satellite data proved useful in reconstructing the fire history of the dump, their spatial resolution was insufficient for detailed monitoring of small-scale thermal anomalies. The study highlights the diagnostic potential of UAV-based remote sensing in post-industrial environments undergoing land degradation but emphasises the importance of field validation for accurate environmental assessment.

**Introduction**

Access to satellite imagery has made environmental monitoring much more straightforward in recent decades (Burke et al., 2021; Chuvieco, 2020; Manfreda et al., 2018). The development of digital infrastructure, including telecommunications networks and databases, enabled rapid processing and broad sharing of this data. A key turning point was the increase in resolution and release of spectral band data, which allowed calculating various indices focused on different aspects of the natural environment to better understand changes occurring in geoecosystems (Zeng et al., 2022). To date, researchers have developed over a hundred vegetation indices based on multispectral imagery (Huang et al., 2021; Xue and Su, 2017), all designed for a proper digital dataset interpretation (Huang et al., 2021). Among them, the Normalised Difference Vegetation Index (NDVI) has become the most commonly used, mainly due to its simplicity, long history of application, and reliance on the accessibility of utilised spectral bands (Eastman et al., 2013; Huang et al., 2021; Pettorelli, 2013; Pettorelli et al., 2011; Tucker et al., 2001). The index is based on the contrast between the highest reflectance in the near-infrared band and absorption in the red band, derived from measurements of optical sunlight reflectance in these wavelengths. As such, NDVI continues to serve as a foundational tool in environmental monitoring, paving the way for developing more specialised indices tailored to specific ecological phenomena.

Beyond general vegetation monitoring, spectral indices have also found wide application in other environmental fields. One of the most prominent examples is fire monitoring, particularly in the context of forest fires (Brey et al., 2018; Cuevas-Gonzalez et al., 2009; Kasischke et al., 1993; Kasischke and French, 1995; Lozano et al., 2010). These indices can be divided into two groups: fire-specific indices (such as Normalized Burn Ratio, Burn Area Index, and Burn Scar Index) and vegetation indices (such as Normalised Difference Vegetation Index, Enhanced Vegetation Index, and Soil Adjusted Vegetation Index). Fire-specific indices are helpful for directly detecting fires and assessing damages (Alcaras et al., 2022; Amroussia et al., 2023; Escuin et al., 2008; Oliveira et al., 2025). Vegetation indices help to monitor the condition of vegetation, which in turn allows for the identification of areas prone to ignition and the development of fire risk maps, as well as assessing the impact and regeneration of vegetation after a fire (Avetisyan et al., 2023; Cuevas-Gonzalez et al., 2009; Lentile et al., 2007; Massetti et al., 2019).

While the widespread use of vegetation indices in fire research typically refers to conventional surface fires – most commonly forest and agricultural fires – it is worth noting that other, less visible types of fires also occur, for which the application of such indices is far less straightforward. There have been various attempts to use environmental indices for detecting subsurface fires, which tend to affect areas like coal seams, peatlands, or landfill sites (Anghelescu and Diaconu, 2024). These fires are challenging to manage due to how hard they are to detect early on and the limited options available for putting them out – often requiring long, complex, and costly interventions. In practice, one of the more effective approaches involves analysing Land Surface Temperature (LST), which helps identify unusual heat patterns at the surface that could point to underground burning (Yuan et al., 2021). These temperature readings are typically derived from satellite data, including sources like MODIS, Landsat, or Sentinel, which provide thermal and short-wave infrared imagery suited for spotting such hidden heat sources (Biswal et al., 2019; Chatterjee, 2006; Kuenzer et al., 2008; Liu et al., 2023; Nádudvari et al., 2021; Saraf et al., 1995; Zhang et al., 2004). However, continuous monitoring of self-heating dumps can reveal the migration, appearance, intensification, or disappearance of these hot spots – especially if they are hot enough to be detected by these sensors (Nádudvari, 2014; Nádudvari et al., 2021). As noted by Biswal et al. (2019), different remote sensing approaches to calculate LST have their advantages and limitations in underground fire detection and mapping and should, therefore, be appropriately selected for specific studies. No matter which method is used, it's essential to avoid the influence of solar radiation, as nighttime or pre-dawn data tend to yield more accurate results compared to daytime data – the presence of sunlight increases the detected heat. Therefore, daytime imagery is not recommended for delineating the extent of coal fires.

One interesting example of subsurface fires is the smouldering of coal-waste dumps, common in active coal mining regions worldwide (Stracher et al., 2016, 2014, 2012, 2010). Their formation is linked to the natural tendency of coal to self-heat and self-ignition (Fabiańska et al., 2019; Gogola et al., 2020; Skotniczy, 2020). However, monitoring these fires proves to be more challenging than tracking large-scale forest or coal seam fires. The main issue lies in the resolution of satellite imagery, which often makes it challenging to detect fire hotspots early enough (Nádudvari et al., 2021; Nádudvari and Ciesielczuk, 2018). These hotspots are typically small, as they cover only a part of the waste dump's surface, and over time, they change in size and/or location. A promising solution to this challenge lies in current technological advancements, particularly using unmanned aerial vehicles (UAVs) equipped with multispectral cameras, which offer resolution up to several centimetres. These devices are increasingly being utilised in post-mining areas by both site managers and researchers around the world (Messinger and Silman, 2016; Ren et al., 2024, 2022; Shao et al., 2023; Thiruchittampalam et al., 2023; Zubíček et al., 2024). So far, their application has primarily focused on thermal monitoring, particularly for rapidly detecting fires and tracking their progression.

The study aims to assess the potential of using environmental indices to monitor subsurface fire changes in coal-waste dumps through multispectral imagery. The research involves the use of UAVs to collect remote sensing data, along with regular on-site field surveys to verify environmental changes. Additionally, the article will evaluate whether these indices apply to this type of fire, primarily based on actual field validation of the environmental and vegetation conditions at the study site.

**Study area**

The study area covered a smouldering coal-waste dump located within the Upper Silesian Coal Basin (USCB) – one of the largest coal mining regions in Europe, located in Poland and the Czech Republic. The investigated site is situated in the city of Chorzów, Poland (Fig. 1), within a densely urbanised and industrialised zone with a long-standing history of mining-related activity dating back to the 19th century (Abramowicz et al., 2025). Initially, the area was a disposal site for metallurgical slag from the Royal Smeltery (org. *Königshütte*). Over time, it was used to deposit coal mining waste from nearby hard coal mines. The dump originally formed a distinctly elevated, conical structure, which underwent multiple morphological transformations before eventually taking its present, relatively flat form. In recent years, the site has been subject to reclamation efforts to create a recreational and park-like landscape. The onset of self-heating and thermal activity within the dump was first documented in the early 2000s (Kotyrba and Siwek, 2017). Since then, the fire has continued to develop, with the fire zone gradually migrating across different parts of the dump. Currently, it appears to be in the final stages of activity. Nevertheless, thermal phenomena remain a source of nuisance for the local population, as unpleasant odours and gas emissions continue to impact their daily lives negatively.

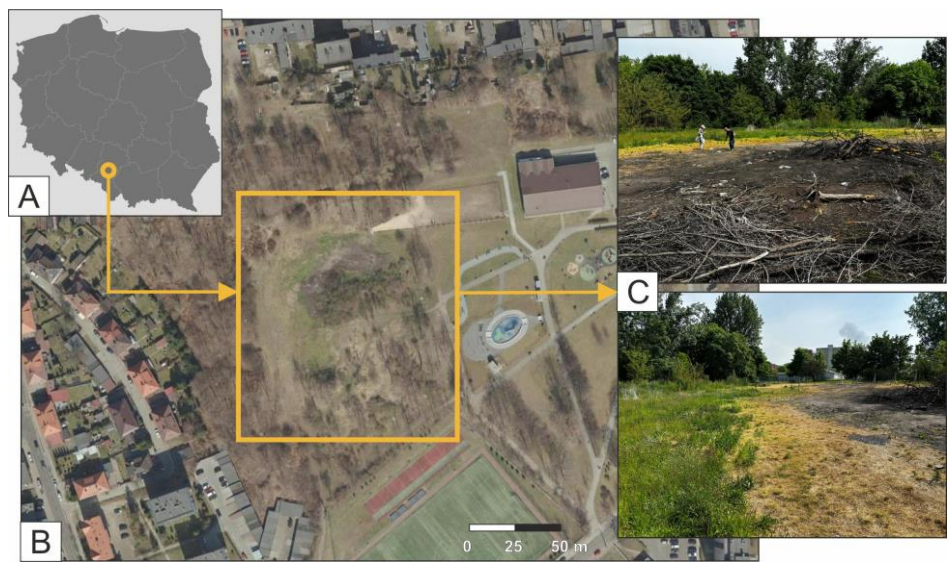

Figure 1. Location of the coal-waste dump in Chorzów: A – on the map of Poland, B – on orthophotomap from 2022 (Google Earth), C – on the field photographs.

**Materials and methods**

The study followed a multi-step workflow combining drone-based field surveys with satellite data to analyse vegetation response over a smouldering coal-waste dump (Fig. 2). The process began with planning and conceptualisation, followed by vegetation mapping during the 2023 growing season, and two UAV surveys in autumn 2023 and spring 2024. In parallel, satellite imagery was sourced from the USGS archive for seasonal comparison. All datasets were processed, georeferenced, and used to calculate vegetation and thermal indices. These formed the basis for identifying spatial trends and assessing the impact of sub-surface fire on vegetation dynamics.

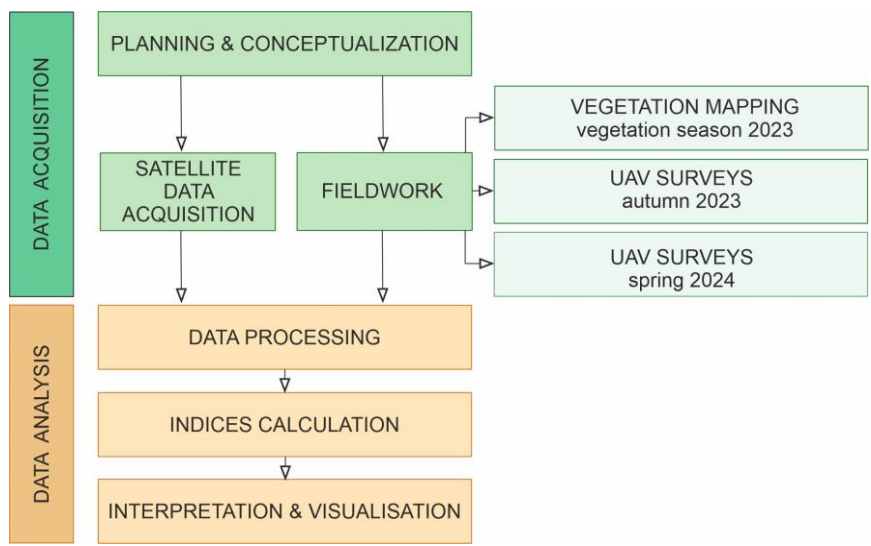

Figure 2. Workflow of data collection and analysis

*Satellite Data Acquisition*

The Landsat images were obtained from USGS (https://earthexplorer.usgs.gov/) and Sentinel data from Copernicus (https://browser.dataspace.copernicus.eu/). The selected datasets cover the study area – burning coal-waste dump and provide a broader temporal and spectral context for analysis. In particular, Sentinel-2 imagery was used due to its high spectral resolution and frequent revisit times. Each scene includes 13 spectral bands, covering visible, near-infrared (NIR), short-wave infrared (SWIR), and red-edge wavelengths (Phiri et al., 2020; Wang and Atkinson, 2018). These bands are especially valuable for vegetation monitoring and detecting thermal and moisture-related anomalies. The spatial resolution of Sentinel-2 ranges from 10 to 60 meters, with key vegetation-related bands such as red and near-infrared (NIR) available at 10-meter resolution, while red-edge bands are provided at 20-meter resolution. The Sentinel-2 data were supplied as Level-1C products (top-of-atmosphere reflectance), which were subsequently processed to bottom-of-atmosphere (surface) reflectance using the Semi-Automatic Classification Plugin (SCP) in QGIS (Congedo, 2021). This step was necessary to ensure spectral consistency across scenes and to enable accurate index calculations.

Landsat ETM+7 imagery, which includes eight spectral bands with a spatial resolution of 30-60 meters (and 15 meters for the panchromatic band), was used in the study to estimate land surface temperature from snow-covered, freezing weather (<0°C), providing valuable thermal information (resolution 60 meters, resampled to 30 m) to complement the high-resolution UAV observations. Additionally, Landsat 4/5 TM data, comprising seven spectral bands with a thermal band at 120-meter resolution (resampled to 30 m), were utilised for historical comparisons. Landsat 8 OLI/TIRS imagery, which includes 11 spectral bands – two of which are thermal infrared bands (TIRS) with a native resolution of 100 meters (resampled to 30 m) – was also employed to enhance the temporal coverage and thermal accuracy of the land surface.

*Field measurements*

Two data collection campaigns were carried out on the selected coal-waste dump in Chorzów – one in the autumn of 2023 and another in the spring of 2024. During each campaign, aerial imagery was captured across multiple spectral bands, including red, green, blue (RGB), near-infrared (NIR), thermal infrared, and red-edge. Data acquisition was conducted using two UAV platforms: a DJI Matrice 210 v2 RTK equipped with an XT2 dual-lens thermal camera (IR + RGB) and a DJI Mavic 3 Multispectral, which features a high-resolution sensor with four dedicated multispectral sensors (green, red, red-edge, and NIR). The days for conducting the measurement campaigns were adjusted to the weather conditions, aiming for windless days without precipitation. Additionally, thermal IR surveys were performed during full cloud cover to minimise the influence of sunlight.

The ground sampling distance for all imagery was approximately 3 cm per pixel for RGB, red-edge and NIR, and 20 cm per pixel for thermal infrared. All collected images were mosaicked and processed using Agisoft Metashape Professional to generate georeferenced orthophotos and analytical layers suitable for further interpretation.

*Vegetation Mapping*

As part of the fieldwork, vegetation mapping was conducted to identify plant species present on the site and to delineate the extent of plant communities covering different parts of the dump. Particular attention was paid to the condition of vegetation in relation to thermal activity zones. Four distinct zones were recognised: the fire spot, the death zone, the transition zone, and the thermally inactive areas (Abramowicz et al., 2021). The health and presence of vegetation were assessed in each of these zones. Additionally, root zone temperature was measured using a 30 cm probe connected to a Raytemp 28 pyrometer, providing ground-level and subsurface thermal data to complement aerial observations. Meteorological conditions were measured using a Trotec BC21 thermohygrometer and a Trotec TA300 anemometer.

*Calculation of Environmental Indices*

The collected remote sensing data were processed and visualised as thematic maps using ArcGIS and QGIS software based on the ETRF2000-PL / CS92 coordinate reference system (EPSG:2180). Raster layers were aligned and prepared for further analysis, and environmental indices were calculated using the raster calculator tool in QGIS. The selected indices were among the most commonly used and were derived from combinations of spectral bands, including red, green, blue, thermal infrared (TIR), near-infrared (NIR), and red-edge. Several vegetation indices were calculated (see Table 1), primarily focusing on NDVI, which provides information about biomass quantity and health. In addition, Green NDVI – a variant more sensitive to chlorophyll content – was used to enhance the detection of subtle vegetation changes. The Enhanced Vegetation Index (EVI) was also included, as it is less affected by dust and smoke, making it particularly useful in fire-impacted areas. Furthermore, the Soil Adjusted Vegetation Index (SAVI) was calculated to account for soil influence in sparsely vegetated zones, using a soil brightness correction factor $L = 0.5$, which is a standard value recommended for areas with moderate to low vegetation cover (Huete, 1988). Finally, the Burn Area Index (BAI) was used to detect regions with low vegetation cover and high levels of charring.

Table 1. Environmental indices used in the study

| Index type | Name of index | Calculation | Reference |
|---|---|---|---|
| vegetation indices | NDVI (Normalised Difference Vegetation Index) | (NIR−Red)/(NIR+Red) | (Eastman et al., 2013; Escuin et al., 2008; Kasischke and French, 1995; Pettorelli, 2013; Pettorelli et al., 2011; Xue and Su, 2017) |
| | GNDVI (Green NDVI) | (NIR−Green)/(NIR+Green) | (Basso et al., 2019; Rahman and Robson, 2016; Sankaran et al., 2018) |
| | GRVI (Green-Red Vegetation Index) | (Green−Red)/(Green+Red) | (Ballester et al., 2019; Motohka et al., 2010; Yin et al., 2022) |

| | SAVI (Soil Adjusted Vegetation Index) | (NIR−Red)/(NIR+Red+L)∗(1+L) | (Gilabert et al., 2002; Huete, 1988; Qi et al., 1994; Rondeaux et al., 1996) |
|---|---|---|---|
| | EVI (Enhanced Vegetation Index) | 2.5∗((NIR−Red)/(NIR+6∗Red−7.5∗Blue+1)) | (Gurung et al., 2009; Matsushita et al., 2007; Villamuelas et al., 2016) |
| fire indices | BAI (Burn Area Index) | $1/((0.1-\text{Red})^2+(0.06-\text{NIR})^2)$ | (Chuvieco et al., 2002; Quintano et al., 2011) |

**Results**

*Historical thermal activity of the coal-waste dump in Chorzów*

Satellite imagery with 30-meter resolution reveals clear signs of surface thermal disturbances in the study area, observed consistently since 1999 in the southern part of the dump (Fig. 3). These surface temperature anomalies are most noticeable during the winter months (January–February) when thermal contrasts are strongest. At that time, clusters of 2 to 5 pixels typically appear warmer than their surroundings by approximately 1°C, indicating localised heat sources likely linked to ongoing weak thermal processes. This pattern can be observed in the most recent imagery as well. However, the location of the thermal hotspot cluster shifts slightly over time by about 1–2 pixels, which is related to geolocation accuracy and minor changes in the spatial dynamics of the subsurface fire.

The surface temperature values extracted from winter scenes clearly reflect the thermal contrast between active and inactive zones. For example, in 1999 (Landsat 7 ETM+), the surface temperature in the thermally active zone reached –3.87 °C, compared to –5.81 °C in inactive areas. A similar contrast is observed in 2004 (–4.23 °C vs. –6.34 °C), 2017 (–3.72 °C vs. –7.20 °C), and 2018 (–2.85 °C vs. –4.42 °C). The anomaly persisted in 2021, with temperature of –6.15 °C in active zones and –8.65 °C in inactive ones. The earliest available scene from 1997 (Landsat 4/5) showed uniformly low surface temperature across the area (–15.89 °C to –14.6 °C), and the most recent scene from 2024 (Landsat 8) also showed a narrow temperature range (–10.95 °C to –10.46 °C). However, in both cases, the limited thermal contrast and apparent homogeneity are likely due to the coarser thermal resolution and reduced sensitivity of these sensors, which are insufficient to detect small-scale thermal anomalies associated with the subsurface fire.

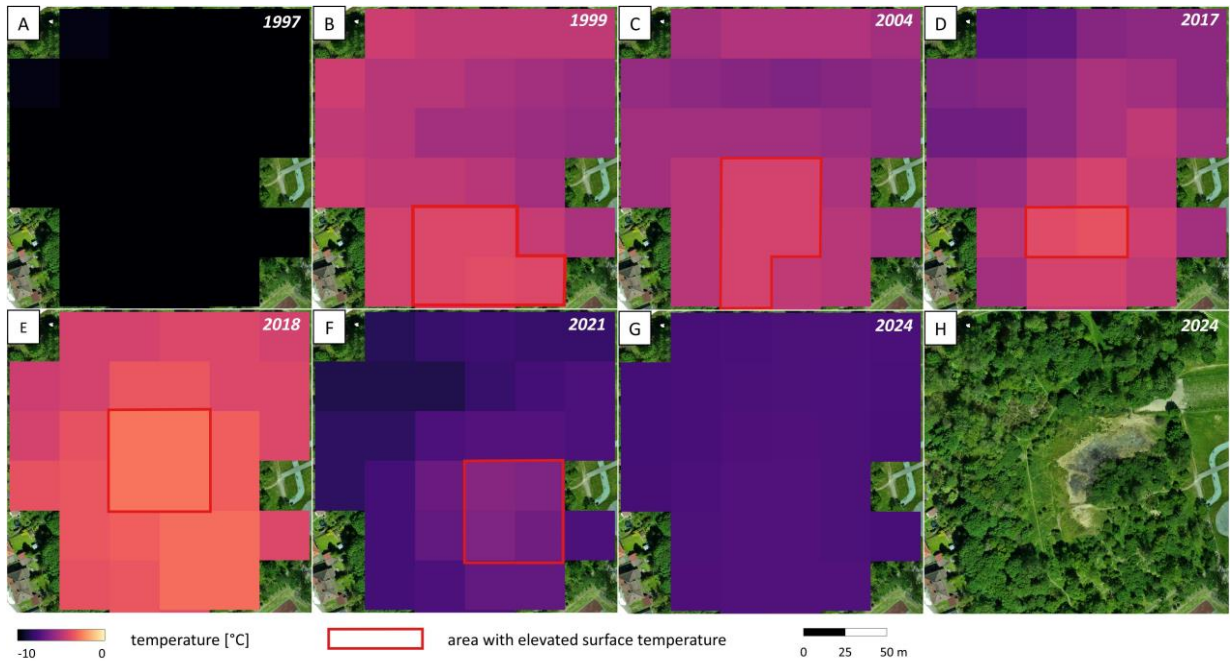

Figure 3. Thermal imagery from 1997-2024 with highlighted temperature anomalies and reference orthophotomap, where: A – 17.12.1997 (Landsat 4/5), B – 31.12.1999 (Landsat 7ETM+), C – 06.03.2004 (Landsat 7ETM+), D – 30.01.2017 (Landsat 7ETM+), E – 09.02.2018 (Landsat 7ETM+), F – 01.02.2021 (Landsat 7ETM+), G – 09.01.2024 (Landsat 8), H – 28.03.2024.

*Thermal characteristics of the coal-waste dump in Chorzów*

Direct point temperature measurements and observations conducted in both autumn and spring clearly indicated substantial contrasts between thermally inactive and active zones, with the latter identified as areas exhibiting elevated ground surface temperatures – typically exceeding background levels – confirmed by both in situ data and thermal remote sensing. In November, surface and subsurface temperature increased despite a relatively low air temperature of –3°C, with 85% relative humidity and moderate winds of 3.3 m/s. A thin but continuous snow cover and light snowfall events likely contributed to insulation effects, allowing the surface to warm up to 18°C and the subsurface layer at -20 cm to reach 78°C in the thermally active zone. By contrast, April was notably dry, with maximum air temperature reaching 14°C, lower humidity (55%), and slightly stronger wind speed (4.1 m/s). The absence of snow and drier conditions allowed for more direct solar heating of the soil surface, resulting in a surface temperature of 27°C, while the subsurface temperature at -20 cm reached 49°C. Although lower than in November, this still indicates a marked capacity for heat retention deeper in the soil, even under dry conditions.

The presence of a subsurface fire was confirmed by remote sensing data, which clearly delineated the extent of the thermally active zone both in November and April, revealing its temporal dynamics (Fig. 4). Since the analysis is based on surface temperature measurements, it reflects the surface expression of the thermal anomaly rather than subsurface conditions, such as those at a depth of –20 cm. The spatial distribution of the zone forms an arc that partially encircles the central part of the coal-waste dump from the north and west. In November, the temperature distribution within the zone was relatively homogeneous, with a recorded maximum surface temperature of 22.2 °C, while temperature in thermally inactive areas ranged from –5 °C to –9 °C. The temperature in April was considerably higher, with inactive areas reaching 6–9 °C and the active zone showing a peak temperature of 27.7 °C. This resulted in a less distinct outline of the fire-affected zone due to the reduced thermal contrast between the reference surfaces and the hotspot. Although the fire zone covered a larger area in April, the thermal anomaly was less homogeneous, with a concentration of peak temperature observed in the northwestern part of the zone.

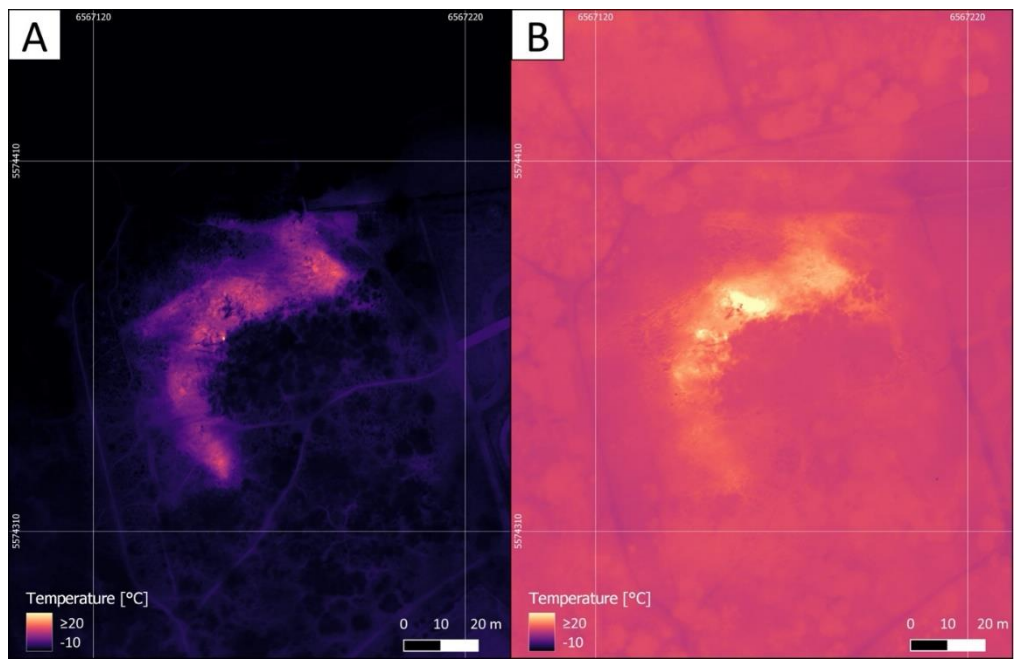

Figure 4. Thermal characteristics of the coal-waste dump during the study period: A – autumn 2023, B – spring 2024 (EPSG:2180)

*Vegetation on the coal-waste dump in Chorzów*

A floristic survey on a coal-waste dump recorded 107 vascular plant species, classified into 39 families and 86 genera. Despite the substantial environmental degradation typical of such post-industrial habitats, the site supports a surprisingly diverse plant assemblage composed of species with a wide range of ecological preferences and syntaxonomic affiliations. Rather than forming distinct, structured communities, many species are linked to broad vegetation classes commonly found in disturbed or human-influenced environments. The dominant non-forest vegetation types are associated with syntaxa such as *Artemisietea vulgaris*, *Molinio-Arrhenatheretea*, *Stellarietea mediae*, and *Rhamno-Prunetea* – all characteristic of habitats shaped by anthropogenic processes. The composition of the flora also reflects clear biogeographical and historical patterns. Apophytes – native species that have adapted to secondary habitats – make up the largest share, accounting for the majority of taxa observed. In terms of life forms, hemicryptophytes are the most frequent, followed by therophytes and megaphanerophytes. Forest species, including both native and non-native trees and shrubs, constitute a significant portion of the assemblage. Non-forest species, on the other hand, are mainly represented by ruderal plants, meadow species, and segetal weeds, most of which are associated with vegetation classes such as *Artemisietea*, *Stellarietea mediae*, and *Polygono-Chenopodietalia*.

The study area is characterised by a patchy arrangement of plant communities shaped by thermal activity (Fig. 5). Central thickets dominated by *Robinia pseudoacacia* and *Lonicera xylosteum* are surrounded by heat-impacted zones. Sparse tufts of *Eragrostis minor* and algal–moss crusts occur near the fire boundary, especially in moisture-retaining depressions. Transitional vegetation includes *Erigeron annuus*, gradually replaced by *Holcus mollis*, with scattered *Plantago lanceolata* and *Phragmites australis*. The site's margins feature trees like *Populus nigra* and *Acer platanoides*, with invasive *Reynoutria japonica* forming a continuous belt. The southern and eastern sections are park-like, with mixed woody vegetation, including *Rosa canina* and *Prunus domestica*.

Vegetation in thermally active areas shows clear disruptions to its natural growth cycle, with some plants flowering and developing outside the typical season while others wither and die during the peak growing period (Fig. 5). Additionally, plant communities tend to form distinct bands around the thermally active zone, suggesting a spatial pattern of colonisation closely linked to the presence of underground fire activity. This is particularly evident in the form of a green band (resembling a fresh meadow) around the thermally active zone in autumn 2023 (Fig. 5A) and as a pale green band overlapping the fire area in spring 2024 (Fig. 5B).

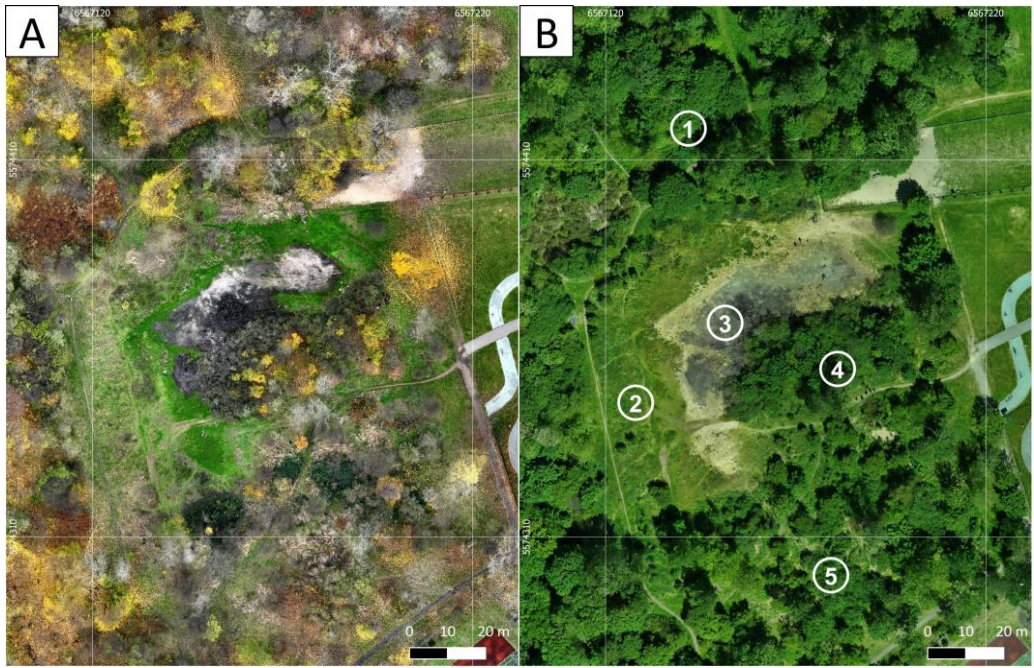

Figure 5. Orthophotomap of the study area: A – autumn 2023, B – spring 2024 with simplified vegetation scheme, where ① tree stands with *Populus nigra, Acer platanoides, Robinia pseudoacacia*, ② community with *Erigeron annuus* and *Eragrostis minor*, ③ fire spot without vegetation, ④ thicket with *Robinia pseudoacacia* and *Lonicera xylosteum*, ⑤ tree stand designed as a park

*Ecological Indices*

The autumn 2023 UAV survey provided detailed insight into the NDVI distribution, revealing index values ranging from -0.09 to +0.91, with a mean of +0.62 and median of +0.66. The highest NDVI values were recorded in areas of dense vegetation along the edges of the burning coal-waste dump, while the lowest values appeared in the northern section of the thermally active zone – specifically within the fire spot – as well as on a nearby sandy dog park (northeastern part of the dump) and paved sidewalks. In the spring dataset, NDVI values ranged from -0.02 to +0.96, with a slightly higher mean of +0.74 and median of +0.85. While the overall distribution remained comparable, this time, the lowest NDVI values extended into the southern part of the thermally active zone as well, suggesting a broader impact of heat stress or substrate limitations on vegetation development.

Ecological indices highlight substantial differences between the two UAV surveys. While the spring survey clearly demonstrated a correlation between areas of low NDVI (below 0.5) and the thermally active zone, the autumn imagery showed a more nuanced picture (Fig. 6). The southern parts of the thermally active area in the fall displayed NDVI values similar to those of the thermally inactive surroundings, indicating possible vegetation regrowth in zones where the surface temperature had moderated.

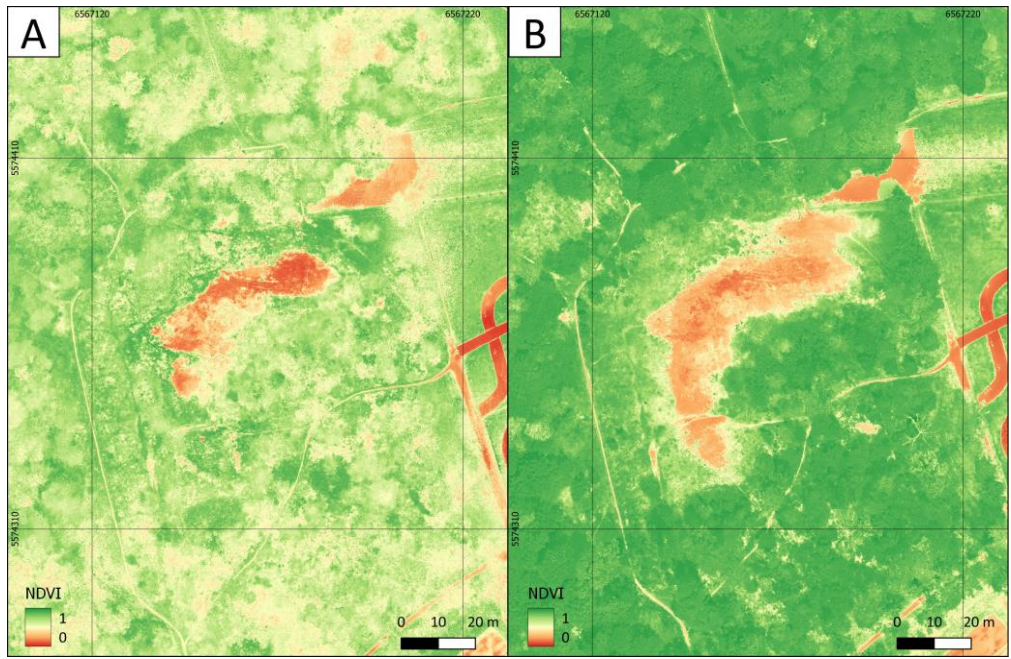

Figure 6. Normalised Difference Vegetation Index of the coal-waste dump during the study period: A – autumn 2023, B – spring 2024

All the ecological indices detected, to varying degrees, anomalies associated with the thermal activity of the coal-waste dump while also subtly reflecting seasonal variability (Fig. 7). Vegetation indices such as NDVI, GNDVI, GRVI, SAVI, and EVI produce largely consistent results – areas affected by fire, where vegetation is absent, consistently show lower index values. While the differences between burned and unburned areas vary slightly depending on the index used, which affects image contrast and visual interpretation, the overall pattern remains the same. In contrast, the Burn Area Index (BAI) presents a distinctly different signal. Burned zones typically show BAI values above 200, distinguishing them from vegetated areas, which exhibit significantly lower values.

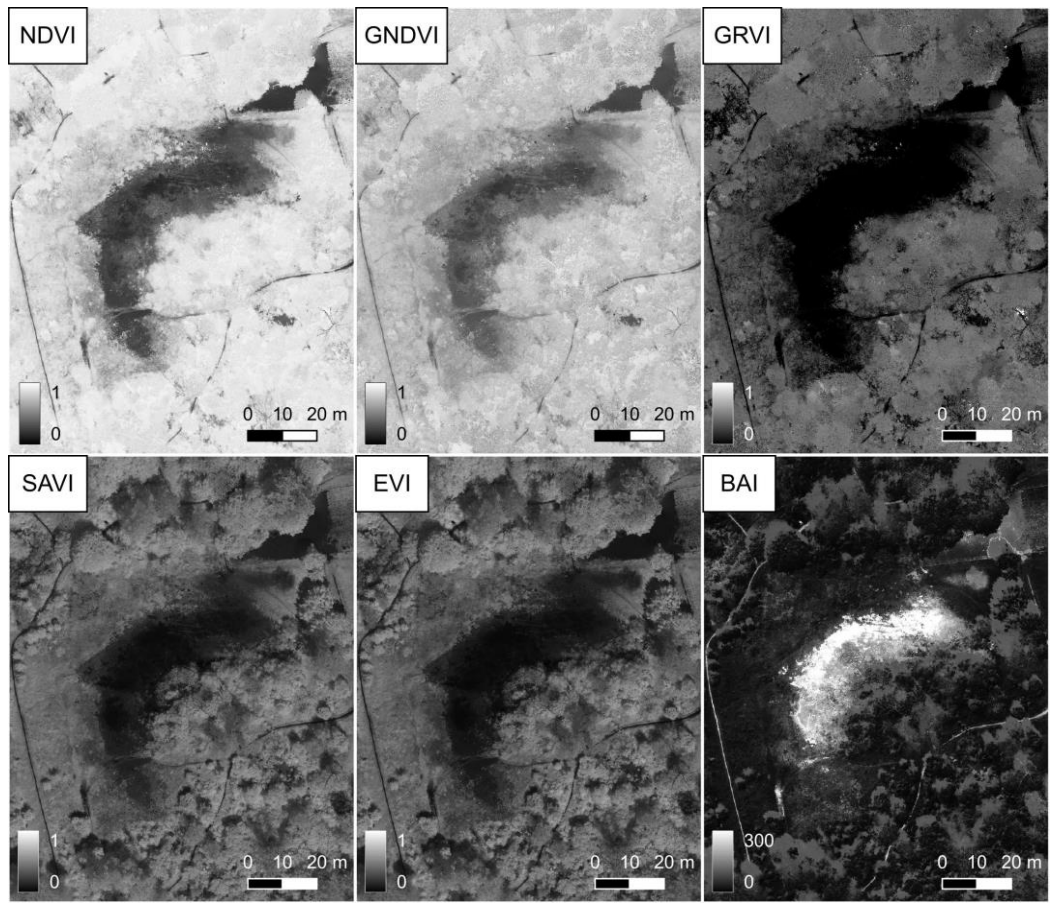

Figure 7. Ecological indices on the coal-waste dump in spring 2024

When comparing the obtained results with satellite imagery data, a clear difference emerges, primarily due to the spatial resolution of the datasets (Fig. 8). Indices calculated from Sentinel-2 imagery show lower pixel values in the burned area. However, the contrast between thermally active and inactive zones is much less pronounced than in data acquired through the UAV survey. This pattern is visible all year round (Fig. 8). During spring, an additional light-toned arc (indicating higher values) surrounds the fire core. This feature corresponds to a band of fresh grassland that forms in these zones during the winter and early spring vegetation periods. In the summer, it disappears due to the air temperature being too high and the low soil humidity. However, elevated vegetation values (high NDVI) can still be observed around the fire during the summer months (due to the leaf cover on trees), while the fire covers the largest area compared to other seasons. In the autumn, vegetation values and fire extent decrease again, reaching their lowest levels in winter. It is especially evident in the January data (Fig. 8, 07.01.2024), where all indices deviate in value from the overall trend due to the snow cover that was present at that time

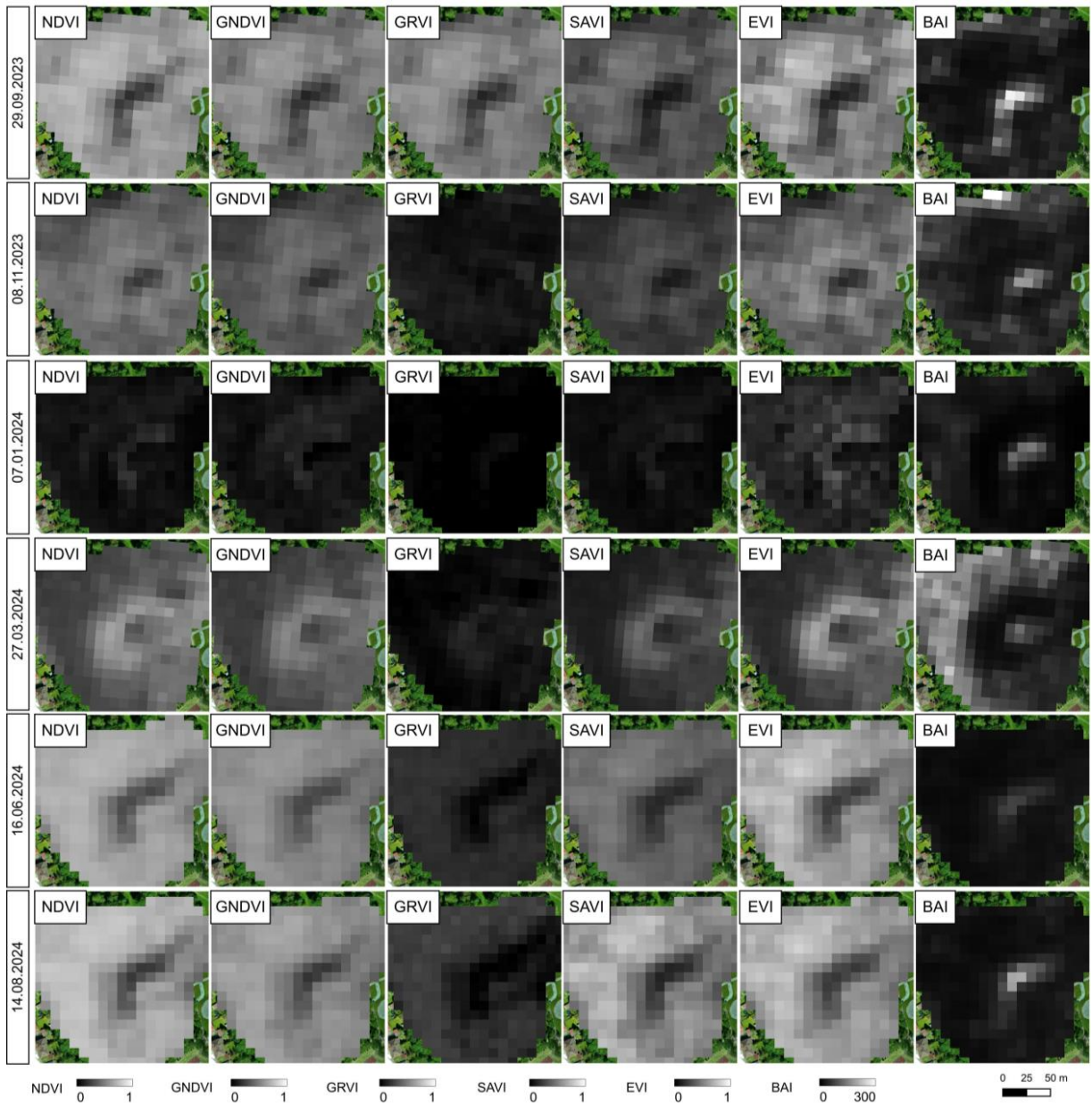

Figure 8. Ecological indices based on Sentinel-2 satellite imagery

In addition to seasonal variability, changes in the spatial distribution of the NDVI index also indicate the gradual movement of the thermally active zone over recent years (Fig. 9). Sentinel-2 imagery from 2015, 2020, and 2024 – captured during the summer months, when vegetation is fully developed – clearly reveals a shift in the area of decreased NDVI values, indicative of fire activity, progressing north-eastwards. This movement suggests that the fire is slowly approaching the recreational section of the park, including areas with outdoor gym facilities and a nearby supermarket. Compared to the lower-resolution infrared images from Landsat (see Fig. 3), Sentinel-2 data offers significantly more detailed insight into the progression of vegetation stress and surface degradation, allowing for more precise tracking of the fire front over time.

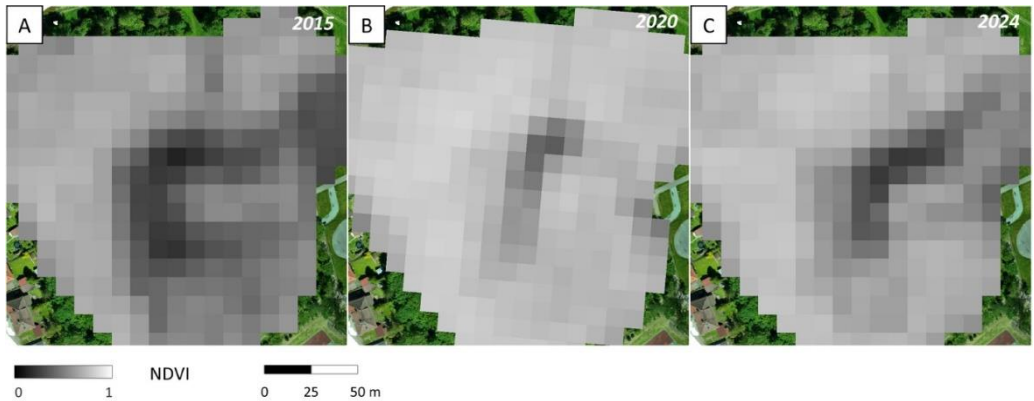

Figure 9. NDVI variability in the period 2015-2024 based on Sentinel-2 satellite imagery: A – 07.08.2015, B – 01.07.2020, C – 14.08.2024

The vegetation cover within the thermally active zone is highly dynamic and varies significantly both spatially and over time, which is clearly illustrated in the cross-sections through the fire spot (Fig. 10). The outer zones (0–20 m and 100–110 m), which are thermally inactive, remain relatively cool and exhibit higher NDVI values compared to the central parts of the fire zone – this is due to the absence of subsurface fire influence. In contrast, the thermally active core shows clear signs of vegetation disturbance, with lower NDVI values caused by the destructive impact of intense subsurface heating. Transitional zones between these extremes are marked by vegetation that appears visually stressed or wilting yet still maintains relatively high NDVI values ranging from 0.5 to 0.7. This paradoxical situation is likely due to residual chlorophyll activity or species-specific spectral responses, which can delay the NDVI drop despite deteriorating plant conditions. Most intriguing, however, is a distinct zone observed in autumn 2023 between the active and inactive areas, where vegetation appears undisturbed (NDVI levels typical of thermally inactive regions), yet elevated surface temperature suggests underlying thermal activity just beneath the surface.

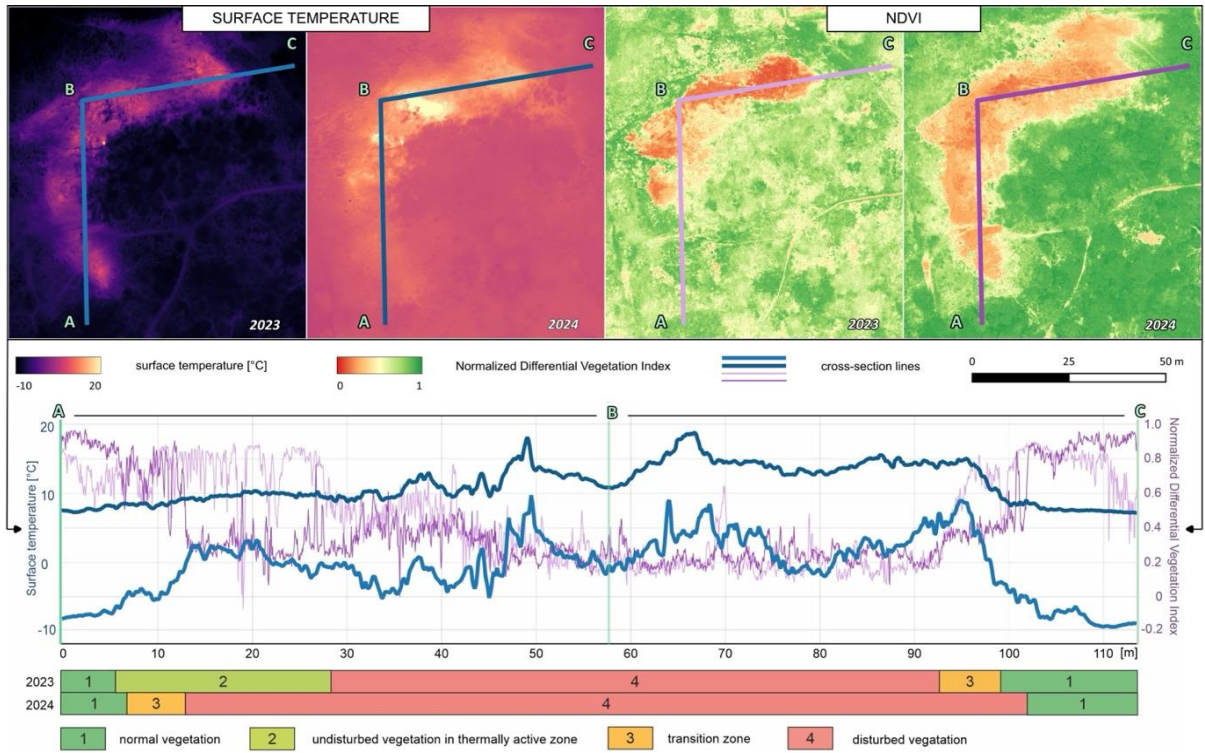

Figure 10. Cross-section through the thermally active zone showing thermal variation and NDVI dynamics

The zonation of vegetation, closely linked to thermal conditions and observed in the field, has also been reported by other researchers studying burning coal-waste dumps worldwide, including those in the Upper Silesian Coal Basin (Abramowicz et al., 2021; Ciesielczuk et al., 2015). However, these zones are rarely thoroughly described or clearly defined. In this context, NDVI proves to be a valuable tool – as it effectively captured the vegetation zonation across areas with differing subsurface temperature within the study site. Five distinct zones can be easily identified based on NDVI values, surface temperature, and plant community composition (Fig. 11).

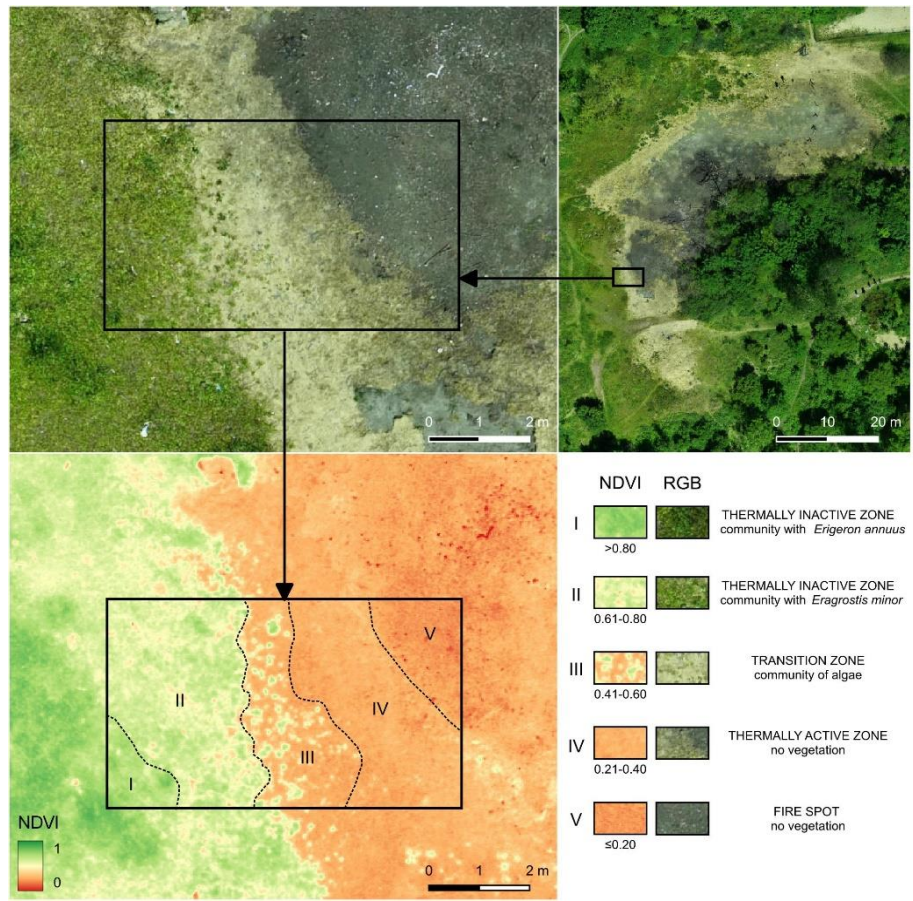

Figure 11. Vegetation zones based on the Normalised Difference Vegetation Index on the burning coal-waste dump in Chorzów

The NDVI seasonal profile in areas affected by subsurface fires differs significantly from the typical pattern observed in undisturbed ecosystems, which has been widely presented in other publications (Dieguez and Paruelo, 2017; Huang et al., 2022; Kaplan and Avdan, 2018; Suzuki et al., 2003; Swanson, 2021). In southern Poland, within the Silesian Upland region – located in a warm transitional temperate climate zone (corresponding to the vegetation zone of deciduous and mixed forests) – the growing season usually extends from the end of March to the beginning of October (Szyga-Pluta et al., 2023). Summer months (June–August) are the period when NDVI values peak, typically reaching 0.7-0.9, as vegetation is fully developed. Around October and November, NDVI values drop sharply to approximately 0.3-0.5 as vegetation prepares for winter. From November to March, NDVI values are minimal – often below 0.2 – because most vegetation in this region is dormant. In the case of subsurface fire areas, the off-season period creates favourable conditions for plant growth within and near thermally active zones (Fig. 12). Elevated temperature in the root zone allows vegetation to develop despite unfavourable air temperature (Ciesielczuk et al., 2015). This results in a situation where NDVI in the surrounding

(non-active) area remains around zero. At the same time, within the thermally active zone, NDVI values become positive and support the growth of particular plant species. During the spring season, NDVI begins to rise in both thermally inactive and active areas, although the increase is more gradual in thermally active zones, where it then declines during the summer. This is mainly due to increased air, surface, and subsurface temperature during summer and the concurrent decrease in soil moisture, which halts vegetation development and leads to dieback. In autumn, the situation reverses – as temperature drops and soil moisture increases, some thermally active zones cool down to levels that promote plant growth, causing NDVI values in parts of these zones to exceed those of their surroundings. This phenomenon is observed, for example, in the southern part of the study area (Fig. 2A, 4A). However, it's important to note that this does not apply to the entire thermally active area. The fire core (the hottest part of the active zone) maintains temperature that are too high for vegetation growth throughout the year, resulting in consistently very low NDVI values year-round.

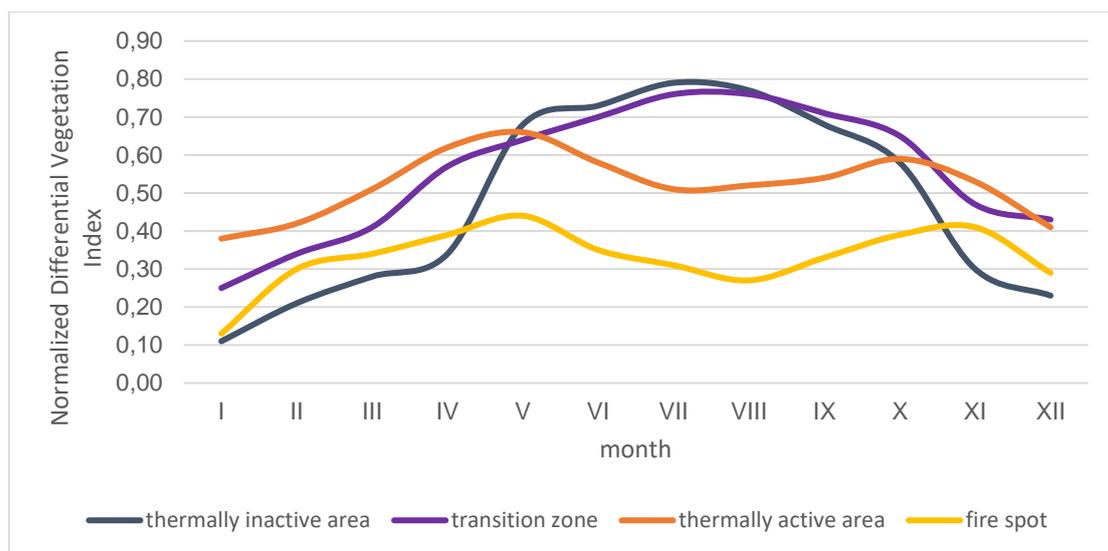

Figure 12. The distribution of Normalised Difference Vegetation Index (NDVI) values throughout the year in thermally active and inactive zones on a burning coal-waste dump (based on Sentinel-2 satellite data in 2024)

**Discussion**

Most studies analysing spectral indices have relied on publicly available satellite imagery, primarily from Landsat ETM+ or Sentinel-2. These datasets are open-access, easy to obtain, offer global coverage, and are updated frequently, which makes them exceptionally convenient and widely used (Burke et al., 2021; Phiri et al., 2020; Wulder et al., 2022). The availability of satellite imagery primarily makes it possible to reconstruct the subsurface fire's development timeline. In this particular study, it allowed us to trace the onset of thermal activity back to the beginning of the Landsat 7 mission, *i.e.* 1999. Although earlier missions, such as Landsat 4 and 5, also provided thermal data, their lower spatial resolution (120 m, resampled to 30 m) limited the ability to detect minor or low-intensity thermal anomalies like those observed at the study site (see Fig. 3). This indicates that thermal processes have been continuously active at the site since at least 1999, persisting for over two decades – a duration commonly observed in long-burning coal-waste fires and similar anthropogenic heat sources (Stracher et al., 2014, 2016). Satellite images also made it possible to determine that the studied dump is not a location known for intense self-heating compared to other dumps in the region. According to the fire classification proposed by Nádudvari et al. (2021), the phenomena observed here fall into the category of weak self-heating for the entire research period (the self-heating intensity index reached values ranging from 0 to 1.5). This could be attributed to the stored material's properties and the waste dump's structural characteristics. It may also suggest that the site is entering the final stages of its thermal activity. Nevertheless, it should be noted that fire, as a dynamic phenomenon, may spread to previously unaffected areas, which could provide new combustible material, which may influence the fire's intensity.

However, the resolution of open-access satellite imagery can still be a limiting factor. While perfectly adequate for large-scale areas or extensive land features, it often falls short when capturing the finer details of smaller objects or localised phenomena. In such cases, UAVs can offer a much more precise alternative, but it also comes with its own set of challenges – higher costs related to owning or renting equipment and hiring certified operators, the need for frequent field visits, and the manual processing of data. There's also the limitation of not having access to historical UAV imagery besides previous own resources. Despite these drawbacks, UAV surveys enable significantly more accurate monitoring and change detection, especially for phenomena that are difficult or even impossible to observe using satellite data (e.g. operations below cloud base). This level of detail can be crucial for better understanding, managing, and predicting geohazards. However, satellite data should always be treated as supplementary materials.

Vegetation indices in burned coal-waste dumps remain anomalously high well into autumn, a pattern that we attribute to locally elevated soil and surface temperature. In effect, the smouldering subsurface fire creates a thermal microclimate similar to a "mini-heat island", sustaining plant activity when surrounding areas are already senescing. This is consistent with experimental and observational studies showing that modest warming accelerates spring growth and postpones fall senescence, lengthening the green season and boosting NDVI. Likewise, urban heat islands can extend plant green-up in cold climates, delaying leaf fall in autumn. By analogy, residual heat from the dump likely prolongs chlorophyll activity and leaf area index of the pioneer vegetation, keeping NDVI

(and related indices like EVI, SAVI or GNDVI) elevated. Case studies of coal-fire and post-mining sites report related phenomena: for instance, Ren et al. (2022) showed that vegetation growth (alfalfa biomass) on a reclaimed coal-waste dump was closely tied to subsurface temperature, suggesting that plants can act as a prior indicator of underground fires. Indeed, thermally active dumps often support distinctive plant assemblages adapted to warm soils (Abramowicz et al., 2021a,b). Crucially, however, this heat-driven greening can mask underlying stress. NDVI values range from –1 to +1, with negative-to-zero values indicating bare or senescent surfaces, so the persistence of high NDVI would usually imply healthy vegetation. However, high greenness under a thermal anomaly does not guarantee ecosystem health. NDVI and similar indices measure chlorophyll and leaf area, not soil quality or pollutant stress, and can remain high even when plants are taking up toxins or suffering moisture deficits. In other words, positive NDVI trends cannot automatically be interpreted as benign; they may reflect climatic forcing rather than improved site conditions (May et al., 2020; Yengoh et al., 2014). In the coal-waste dump in Chorzów, the fire's heat probably keeps plants green and delays autumn browning, but it may also conceal mining legacies (contaminants, altered nutrients or hydrology) that would otherwise become apparent. Thus, the anomalously high greenness on the coal-waste dump – driven by residual fire heat – could lead remote assessments to underestimate ecological stress, underscoring the need for caution when using VIs alone to monitor reclamation success (Yengoh et al., 2014).

Although the primary focus of this study was on vegetation indices, the Burn Area Index (BAI) was also included to capture better the spatial extent and effects of subsurface thermal activity. While indices such as the Normalized Burn Ratio (NBR) and its derivative dNBR are commonly used in fire-related studies (Escuin et al., 2008), they weren't applied here. The main reason is that the UAV-based imagery used in this research did not include the SWIR band, which is essential for calculating NBR. Since dNBR is based on changes in NBR over time, it was also excluded. In theory, both indices could have been derived from Sentinel-2 satellite data; however, the SWIR band in Sentinel-2 has a spatial resolution of 20 meters, which is lower than the 10-meter resolution maintained throughout this study. Using such data would have disrupted the spatial consistency of the analysis and prevented a direct comparison with UAV-based results – an essential aspect of the methodology. Given the relatively small size of the fire-affected area, high-resolution UAV data were critical to capturing spatial patterns accurately. Satellite imagery (e.g. Sentinel-2), with its lower spatial resolution, would likely have overlooked important details. Therefore, BAI was selected as a suitable alternative – it can be calculated from both UAV and satellite imagery using bands available at 10-meter resolution, and it offers a reliable means of identifying fire-related surface changes.

**Conclusions**

In thermally active zones of burning coal-waste dumps, NDVI exhibits seasonal variability, but it differs significantly from the patterns observed in areas unaffected by subsurface heating. This alternative seasonality is marked by the onset of a vegetation period during the meteorological winter, specifically in areas where the root-zone temperature is elevated but not extreme.

NDVI analyses at a fine spatial scale are essential for accurately assessing environmental conditions. They can effectively complement broader-scale remote sensing efforts, especially when paired with field verification of key areas. Without contextual knowledge of site-specific factors such as vegetation types, terrain forms, and land-use history, remote sensing data alone may lead to misleading interpretations.

Vegetation dynamics in burning areas differ fundamentally from those in the surrounding landscape. These changes are driven by altered soil moisture and thermal regimes, which can either accelerate or delay vegetation development, producing a localised shift in topoclimatic conditions.

In the case of small subsurface fire zones, the current spatial resolution of satellite-based spectral imagery remains insufficient for capturing detailed environmental characteristics through vegetation indices. High-resolution UAV data, therefore, play a crucial role in identifying and monitoring such localised phenomena.

**Acknowledgments**

This research was funded in whole by National Science Centre, Poland [Grant number: 2023/07/X/ST10/00540].